      [2001/06/19 v1.4g Il Nuovo Cimento]
\documentclass{cimento}

\usepackage{graphicx}  
\title{Search for the host galaxy of GRB~050904 at $z=6.3$}
\author{K.~Aoki\from{ins:s}\ETC,
H.~Furusawa\from{ins:s},
K.~Ohta\from{ins:k},
T.~Yamada\from{ins:s}\from{ins:o}
        \atque
N.~Kawai\from{ins:t}}
\instlist{
\inst{ins:s} Subaru Telescope, Hilo, Hawaii, U.S.A.
\inst{ins:k} Department of Astronomy, Kyoto University, Kyoto, Japan
\inst{ins:o} National Astronomical Observatory of Japan, Mitaka, Tokyo, Japan
\inst{ins:t} Department of Physics, Tokyo Institute of Technology, Tokyo, Japan
}
\PACSes{
\PACSit{98.70.Rz}{gamma-ray sources; gamma-ray bursts}
\PACSit{98.62.-g}{Characteristics and properties of external galaxies and extragalactic objects}
\PACSit{98.62.Ai}{Origin, formation, evolution, age, and star formation}
}
\begin{document}

\maketitle

\begin{abstract}
We present the results of deep imaging of the field of GRB~050904
with Suprime-Cam on the Subaru 8.2m telescope.
We have obtained a narrow-band (130 \AA) image centered at 9200 \AA~(NB921)
and an $i'$-band image with total integration times of 56700 and 24060 s, 
respectively.
The host galaxy was not detected within $1''$ of the afterglow position.
An object was found at $1.5''$ NE from the position of the afterglow,
but clear detection of this object in the $i'$-band image rules out
its association with the burst.
We obtained a limit of $> 26.4$ AB magnitude ($2''$ diameter, 3 $\sigma$)
in the NB921 image for the host galaxy, corresponding to 
a flux of $6.0 \times 10^{28}$ erg/s/Hz at rest 1500 \AA~assuming a flat spectrum
of the host galaxy.
The star formation rate should be less than 7.5 (M$_{\odot}$/yr) based on the
conversion rate by \cite{ref:madau}. 
This upper limit for the host of GRB 050904 is consistent with the
star formation rate of other gamma-ray burst host galaxies around redshift of 2 or
less. 
\end{abstract}

\section{Deep imaging of the field of GRB~050904}
GRB~050904 was triggered on 2005 September 4, and found to be 
the highest redshift ($z=6.295$) gamma-ray burst (GRB) known to date 
\cite{ref:kawai}.
In order to search for the host galaxy of GRB~050904,
we carried out imaging of the field of GRB~050904 with Suprime-Cam 
on the Subaru 8.2-m telescope
under the photometric conditions on 2005 December 28 - 2006
January 3 (UT), when it was 115 days after the burst.
We used a narrow-band (130 \AA) filter centered at 9200 \AA~(hreafter, NB921)
and the $i'$-band filter.
Although NB921 has narrow width, it has two merits.
It does not include Ly break and covers only the continuum light of an object at $z=6.3$.
The sky background at 9200 \AA~ is lower than the surrounding region because 
OH sky emission lines are weaker at that integration.
The $i'$-band images were taken to discriminate foreground galaxies.
The integration times were 56700 s in the NB921, and 24060 s in the $i'$-band.

\section{Results}
\begin{figure}
\begin{center}
\includegraphics[width=5cm,clip,keepaspectratio]{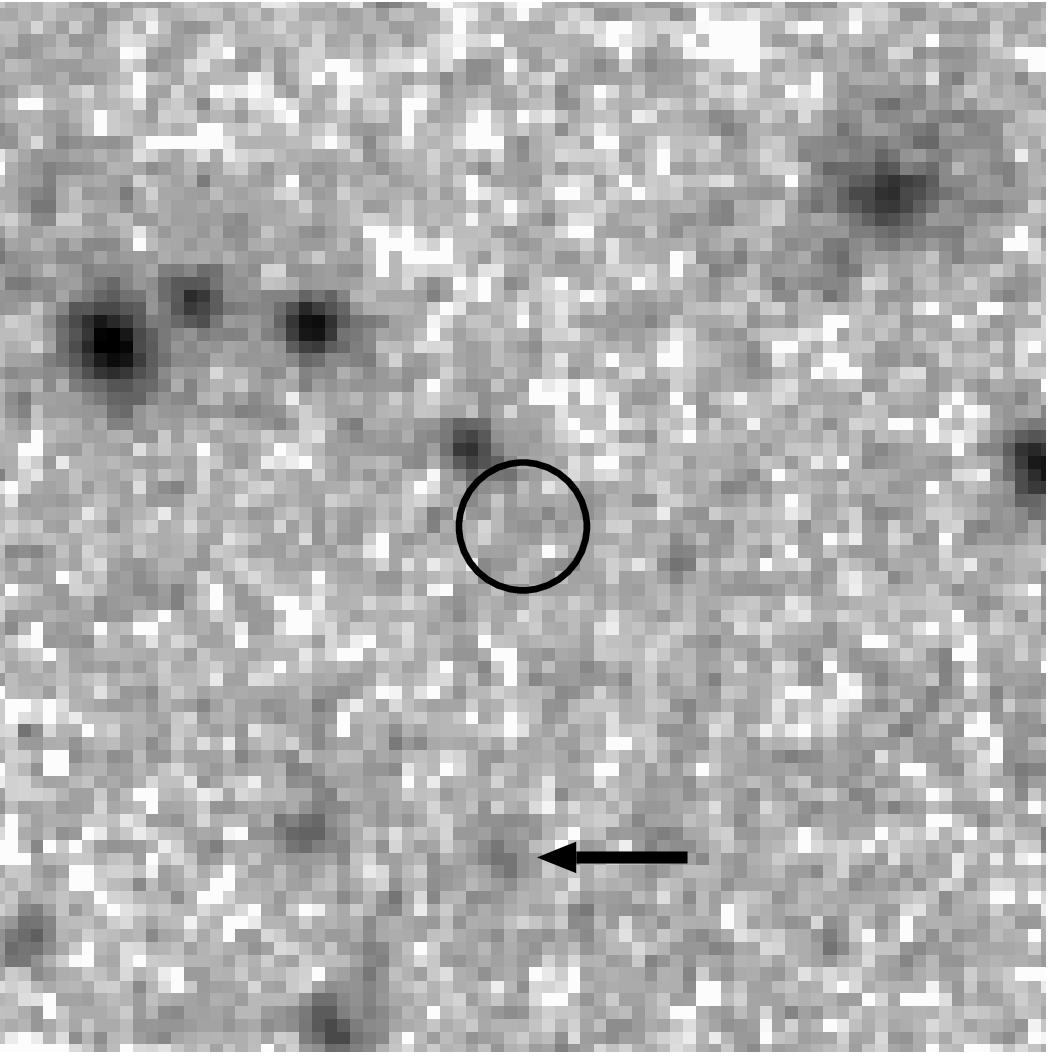}     
\includegraphics[width=5cm,clip,keepaspectratio]{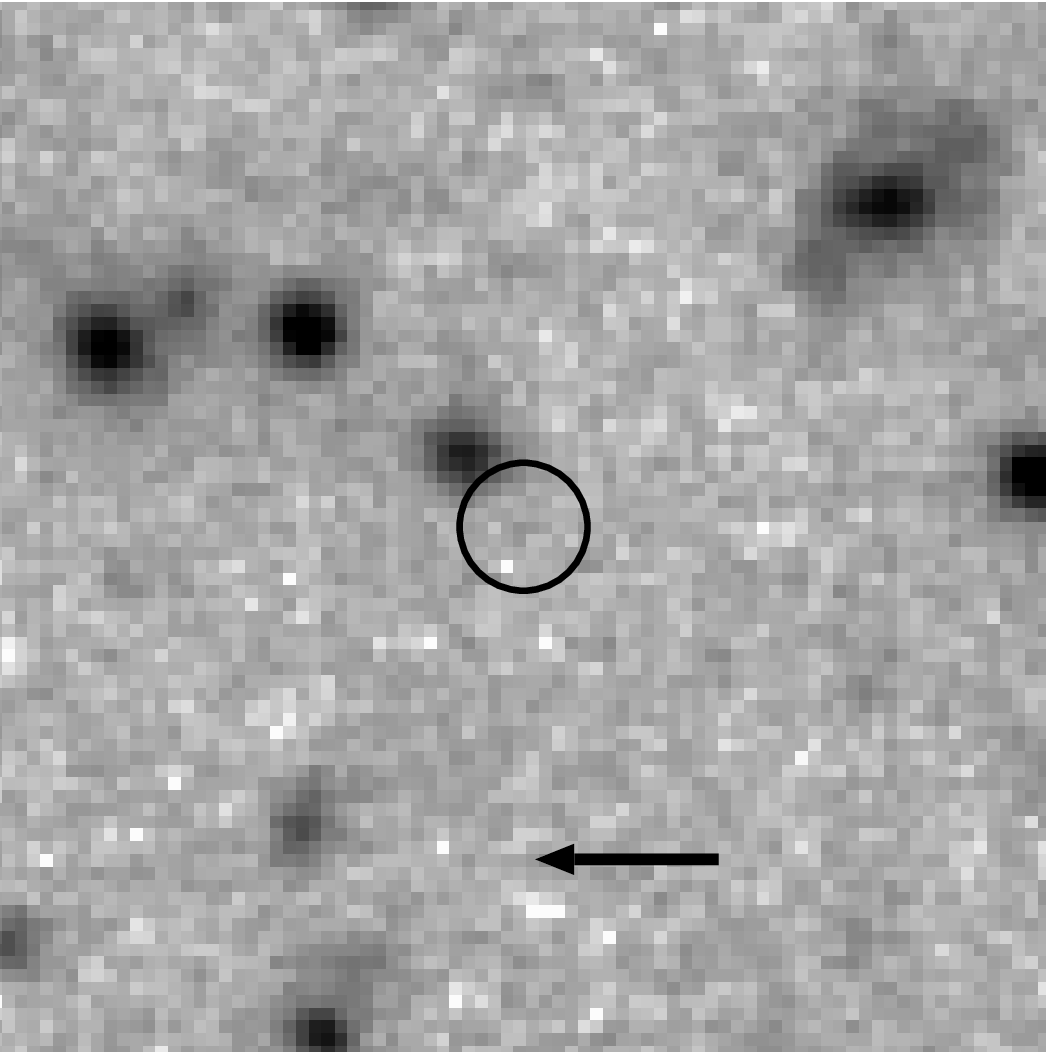}     
\caption{Images of the field of GRB 050904. 
The north is up, east is left. The field of view is $16'' \times 16''$.
The circle indicates the GRB position and its diameter is $2''$.  
The arrow indicates an object located $5''$ south of the GRB.
({\it Left}) The NB921 image. ({\it Right}) The $i'$-band image.
}
\end{center}
\end{figure}
The NB921 and $i'$-band images of the field of GRB~050904 ($16'' \times 16''$)
are shown in Figure 1.
The limiting magnitudes ($2'' \phi$, 3$\sigma$ ) are 
 26.4 AB mag. in the NB921, and 27.2 AB mag. in the $i'$-band.
No object ($> 3\sigma$) is found at the position of the GRB.
The nearest object located at $1.5''$ NE is clearly detected at both the NB921 and $i'$-band image. 
This fact means that no Ly break exists in the objects's spectrum at 8500 \AA, 
and that its redshift is smaller than 6.
\par
The limiting magnitude of 26.4 mag at 9210 \AA~corresponds to the luminosity of $6.0 \times 10^{28}$ erg/s/Hz at 1260 \AA~assuming $H_{0}=70$ km/s/Mpc, $\Omega_{m}=0.3$, $\Omega_{\Lambda}=0.7$.
Assuming a flat spectrum of the host galaxy, we used the following relation 
\cite{ref:madau} to estimate the star formation rate:
$$ SFR(M_{\odot}/yr) = L_{UV} (erg/s/Hz)/ 8.0\times10^{27}. $$
The upper limit of the star formation rate is 7.5 M$_{\odot}$/year.  
This value is consistent with the one estimated by \cite{ref:berger} using $HST$ data.
This moderate star formation rate is similar to those of lower redshift GRB host galaxies~\cite{ref:chris}.
\par
In the NB921 image an object ($2.5\sigma$, 26.6 mag) is located at $5''$ south 
of the GRB. 
This object is also detected in the $HST$ $H$-band image~\cite{ref:berger},
however, it is not detected in the $i'$-band image.
These facts suggest that its possible redshift is larger than 6.
The separation of $5''$ between the object and the GRB corresponds to 28 kpc 
($H_{0}=70$ km/s/Mpc, $\Omega_{m}=0.3$, $\Omega_{\Lambda}=0.7$). 
Threfore, it may reside in the same dark halo where the GRB occurred.


\begin{thebibliography}{0}
\bibitem{ref:kawai} \BY{Kawai \etal} \IN{Nature}{440}{2006}{184}
\bibitem{ref:madau} \BY{Madau, Pozzetti, \atque Dickinson} \IN{ApJ}{498}{1998}{106}
\bibitem{ref:berger} \BY{Berger \etal} astro-ph/0606689
\bibitem{ref:chris} \BY{Christensen, Hjorth, \atque Gorosabel} \IN{A\&A}{425}{2004}{913}
\end{thebibliography}
\end{document}